\documentstyle[epsfig]{aipproc}
\begin{document}
\def\etal{{\it et al.\ }}
\def\hmpc{{h^{-1}}{\rm Mpc}}

\title{Cosmological Implications of Lyman-Break Galaxy Clustering}
\author{Risa H. Wechsler$^1$, Michael A. K. Gross$^2$,
Joel R. Primack$^1$, \\George R. Blumenthal$^3$, Avishai Dekel$^4$, and 
Rachel S. Somerville$^4$}
\address{$^1$Physics Department, University of California, 
Santa Cruz, CA 95064\\
$^2$NASA/Goddard Space Flight Center, Greenbelt, MD 20771 \\
$^3$Astronomy \& Astrophysics Department, University of California, 
Santa Cruz, CA 95064\\
$^{4}$Racah Institute of Physics, The Hebrew University, 
Jerusalem, 91904, Israel}
\maketitle

\begin{abstract}
We review our analysis of the clustering properties of
``Lyman-break'' galaxies (LBGs) at redshift 
$z \sim 3$, previously discussed in Wechsler \etal \cite{w98}.  
We examine the likelihood of spikes found by Steidel \etal \cite{s98}
in the redshift distribution of LBGs, within a suite of models for the 
evolution of structure in the Universe. 
Using high-resolution dissipationless N-body simulations, we analyze deep 
pencil-beam surveys from these models in the same way that they are actually 
observed, identifying LBGs with the most massive dark matter halos. 
We find that all the models (with SCDM as a marginal
exception) have a substantial probability of producing spikes similar to those 
observed, because the massive halos are much more clumped than the underlying 
matter -- i.e., they are biased.  Therefore, the likelihood of such a spike is 
not a good discriminator among these models. The LBG correlation functions are 
less steep than galaxies today ($\gamma \sim 1.4$), 
but show similar or slightly 
longer correlation lengths.  We have extened this analysis and include a 
preliminary comparison to the new data presented in Adelberger \etal 
\cite{adel}.
We also discuss work in progress, in which we use semi-analytic models to 
identify Lyman-break galaxies within dark-matter halos.
\end{abstract}

\section*{Introduction}
Recently, Steidel \etal (S98) \cite{s98} discovered
that the redshift distribution of ``Lyman-break'' galaxies (LBGs) in a
pencil beam reveals a large ``spike'' in the LBG
distribution near $z\simeq 3.1$. This spike corresponds to a
fractional overdensity of LBGs of a few hundred percent over a
comoving scale of order $\sim 10-20\hmpc$.  Since then, they have compiled 
a sample of more than 600 galaxies in six fields, with measured 
redshifts between about 2.5 and 3.5 \cite{adel}, \cite{s98b}. 

At a first glance, such a high peak seems
surprising, since it suggests substantial nonlinear clustering on
rather large scales.  In fact, we \cite{w98} and other authors 
\cite{bagla}, \cite{jing}, \cite{governato} find from simulations
that these spikes
arise naturally in a variety of cosmologies, provided that there is substantial
galaxy biasing.

Here we compare the clustering properties of the LBGs to those expected in 
high-resolution simulations of four different cosmological models, by 
identifying $z\sim3$ LBGs with the most massive halos in our simulations at 
that redshift.  We are currently working on improving this analysis,
both by comparing to the increasing amounts of new data, and by 
using semi-analytic models to explore different ways of identifying LBGs 
with halos in N-body simulations.

\section*{How probable are the Spikes?}
Our methods are described fully in Wechsler \etal (W98) \cite{w98}, and will
be described only briefly here.  The data is taken in $9'$x$9'$ fields; the
redshift distribution is binned in bins of $z=0.4$.  In determining the
probability of spikes, our statistics consider each pixel separately, where a
pixel is one z=0.4 redshift bin by the size of an angular field: $9'$x$18'$ (in the case of 
data from S98), or $9'$x$9'$, (in the case of the newer data from Adelberger 
\etal (A98)
\cite{adel}).  The number of galaxies in each pixel 
is divided by its 
selection function; we then consider the galaxy overdensity per pixel:
$\delta_g=(N-\bar{N})/\bar{N}$.  These data are then compared to 
N-body simulations by Gross \etal 
\cite{grossetal} 
of four cosmological models: SCDM ($\Omega_0 = 1, h = .5, \sigma_8 = 0.67$), 
CHDM ($\Omega_0 = 1, \Omega_{cdm} = 0.8, \Omega_\nu = 0.2, h = .5, \sigma_8 = 0.72$),
OCDM ($\Omega_0 = 0.5, h = .6, \sigma_8 = 0.66$),
$\Lambda$CDM ($\Omega_0 = 0.4, \Omega_\Lambda = 0.6, h = .6, \sigma_8 =0.77$),
in a 75 $\hmpc$ box; 57 million CDM particles (+ 113 million HDM particles in
CHDM).
The halos are identified as virialized regions at $z \sim 3$.
To assign LBGs to halos we assume that one LBG resides in each massive halo,
then choose a mass cutoff to match the observed number density of candidates
and randomly select 40\% of these (because S98 find redshifts for $\sim$ 40\% 
of candidates). The mass cutoff ranges from $3\times10^{11}$ for CHDM to 
$9\times10^{11}$ for SCDM.   We then ``observe'' pixels the size of the 
observational determined pixels, and compared the statistic 
$\delta_g$ to that calculated from the data.  
The cumulative probability distribution of this
galaxy overdensity statistic is plotted in Figure 1, 
both for the analysis 
done in W98, which compared only to the data from the  
one $9'$x$18'$ field published by S98, and compared to
the more complete data set from six $9'$x$9'$ fields published by 
A98.  
We find that the probability of observing a spike as large as 
the largest seen by S98 (corresponding to $\delta_g=2.6$) is about 
37\%, 31\%, 27\%, and 
6\% for CHDM, OCDM, $\Lambda$CDM, and SCDM, respectively.  The reason for the
small probability for SCDM seems to be due to the shape of the power spectrum
\cite{w98}, \cite{adel}.  The distribution of the newer data set published in
A98 looks relatively well fit by all of the models; a full analysis will be
published elsewhere.  In future work, we will also take into better account 
a more accurate selection function for the data.

\begin{figure}[t] 
\hskip 1pc
\begin{minipage}[b]{.47\linewidth}
\centerline{\epsfig{file=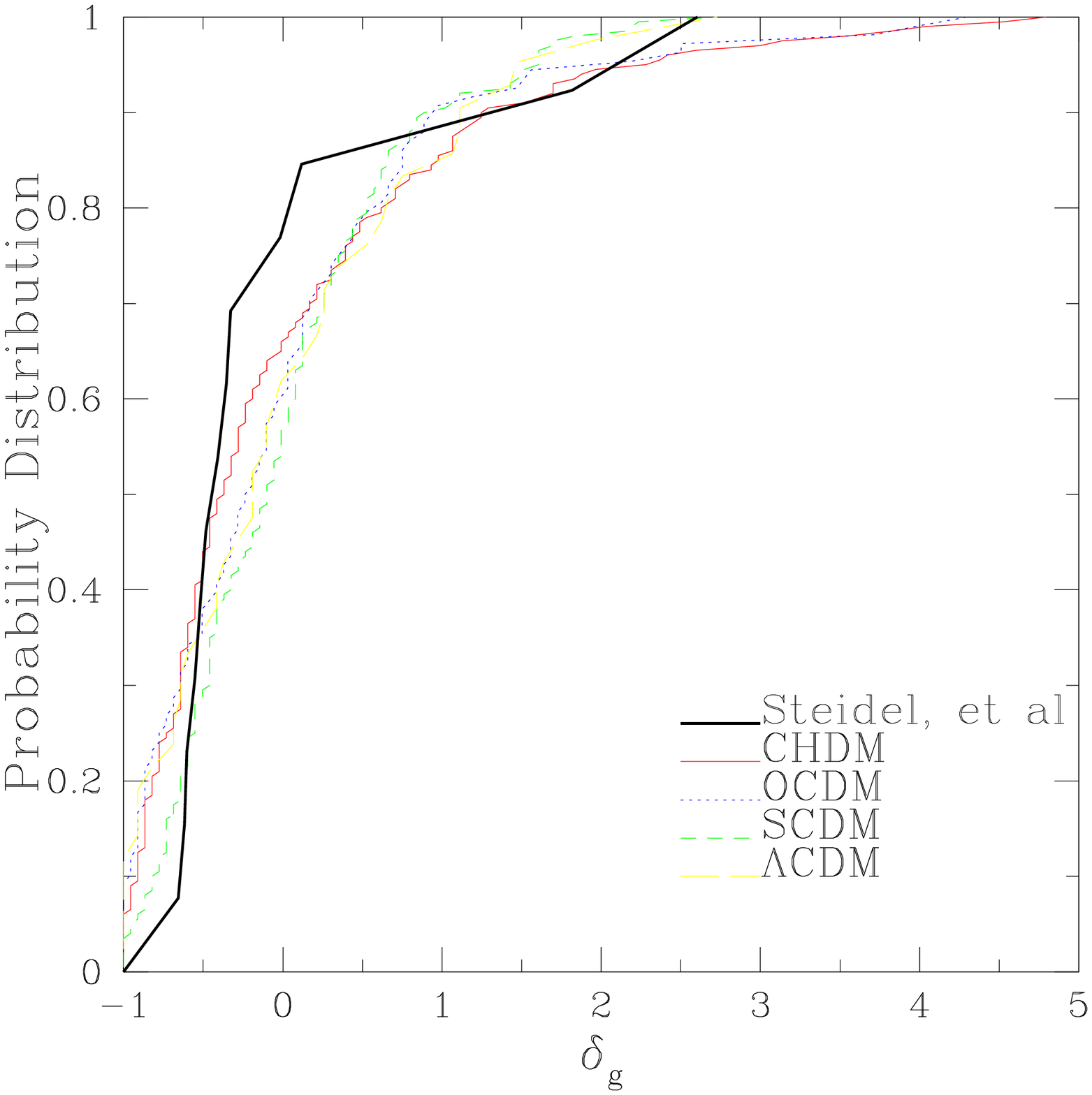,width=\linewidth}}
\end{minipage}\hfill
\begin{minipage}[b]{.47\linewidth}
\centerline{\epsfig{file=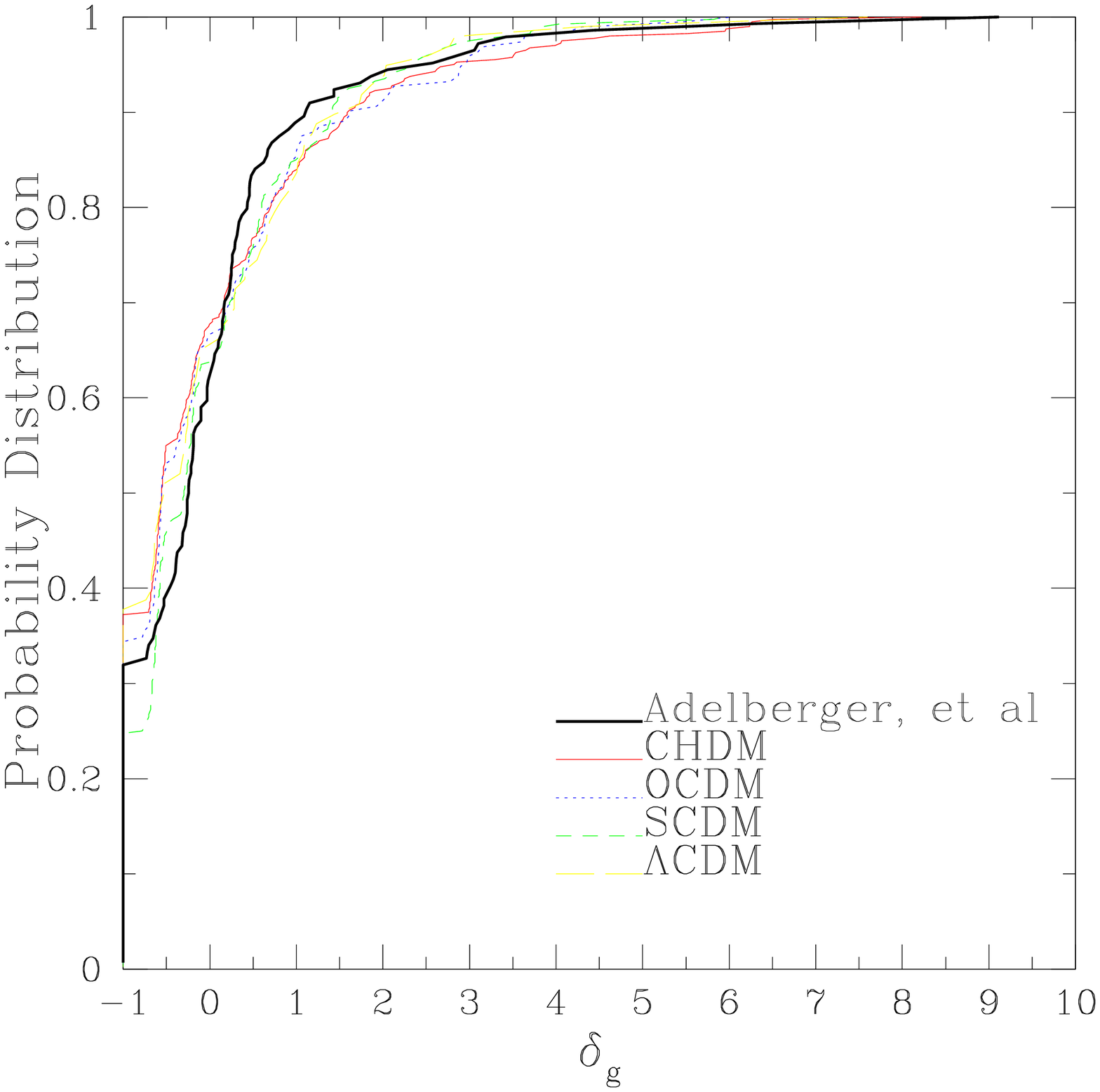,width=\linewidth}}
\end{minipage}\hfill
\vspace{10pt}
\caption{
(a) The cumulative probability distribution of 
$\delta_{\rm g}$ for the S98 data as well as for the 
massive halo distribution in the four models considered here. The distribution 
reflects the relative excess of halos in each pixel.  Spikes the size of the
largest spike found by S98 are found with reasonable probability 
($\sim25-40\%$) in all models except SCDM, and with marginal probability in 
SCDM.
(b) The cumulative probability distribution of  
$\delta_{\rm g}$ for the newer data set of A98 as well 
as for the distribution of massive halos in the four cosmologies considered. 
The
distribution seems reasonably well matched by all of the models.}
\end{figure}

\section*{Bias \& Correlation Function}
The clustering of LBGs is found to be biased with respect to the
underlying dark matter.  The bias factor $b$, 
defined as $b=\delta_g/\delta_m$, 
has an average value of about 2-5 depending on the cosmological model.
High-density regions are the most biased.
\\
We have measured the correlation function for the LBGs in our 
simulations.  The best fit parameters are: $r_0$ [$h^{-1}$ Mpc] = 
3.3, 5.1, 5.0, 7.3, and $\gamma$ = 1.7, 1.6, 1.6, 1.5, for SCDM,
CHDM, OCDM, and $\Lambda$CDM, respectively.
A98 have calculated the correlation function using a
counts-in-cells method, and find correlation lengths of 
$\simeq 4 \pm 1, 5 \pm 1$, and $6 \pm 1 (h^{-1} $Mpc) for $\Omega_M$ = 1, 0.2 
open, and 0.3 flat, assuming that $\gamma$ = 1.8 
(cf. \cite{giavalisco}).

\section*{Adding Semi-Analytic Models}
In previous work, we identified LBGs by making the assumption that one 
object resides in each massive halo above some mass cutoff.  Alternatively, 
we can use semi-analytic models \cite{sp98}, \cite{spf98}, \cite{psfw} 
to predict the location of objects luminous enough to be observed as LBGs --
they may reside in less massive halos, or there may be more than one per halo.
We are currently exploring how different models of galaxy formation
affect the clustering properties of the Lyman-break galaxies when identified 
in this way -- which may help to distinguish between galaxy formation models 
and to determine the nature of the Lyman-break galaxies.

\section*{Conclusions}
Large peaks in the observed redshift distribution of LBGs are common
at $z \sim 3$ in several cosmological models.  Galaxy formation is biased 
at high-z in all cosmologies we have considered, with a bias factor of 
$\sim 2-5$. The clustering properties will probably not distinguish 
between different cosmologies, without independent information about 
the bias.  We find a similar correlation length but slightly shallower 
correlation function slope compared with that observed.
We find that the bias and clustering properties are more strongly affected by 
the shape of the power spectrum than by the mass density; a model 
with a shallow power spectrum, like SCDM, seems to be somewhat less
clustered than the data.


A preliminary analysis of the data published by A98 shows that
the original spike \cite{s98} is one of the highest in all surveyed fields, 
and the new
data is in fairly good agreement with all four cosmological models. Major 
improvements to our prior analysis can be made both by comparing to the 
still-increasing amounts of new data, and by improving our method of assigning
LBGs to dark-matter halos.  By using semi-analytic models to test and improve
upon our method of identifying LBGs within halos from N-body simulations, 
we may be better able to understand the clustering properties and nature of the
Lyman-break galaxies, and may begin to distinguish between galaxy formation 
scenarios.

\end{document}